\documentclass[conference]{IEEEtran}
\usepackage{cite}
\usepackage{amsmath,amssymb,amsfonts}
\usepackage{graphicx}
\usepackage{subcaption}
\usepackage{pgfplots}
\pgfplotsset{compat=newest}
\usepackage{tikz}
\usepackage{svg}
\usepackage{textcomp}
\usepackage{xcolor}
\usepackage{hyperref}
\usepackage{algorithm}
\usepackage{algpseudocode}
\usepackage{listings}
\usepackage{multirow}
\usepackage{cleveref}
\def\BibTeX{{\rm B\kern-.05em{\sc i\kern-.025em b}\kern-.08em
    T\kern-.1667em\lower.7ex\hbox{E}\kern-.125emX}}
    
\begin{document}

\title{Improving Residential Safety by Multiple Sensors on Multiple Nodes for Joint Emergency Detection}

\author{
	\IEEEauthorblockN{
		Artur Sterz\IEEEauthorrefmark{1},
		Markus Sommer\IEEEauthorrefmark{1},
		Kevin Lüttge\IEEEauthorrefmark{1},
        Bernd Freisleben\IEEEauthorrefmark{1}
	}\\
	\IEEEauthorblockA{
		\IEEEauthorrefmark{1}\textit{Dept. of Mathematics \& Computer Science, University of Marburg, Germany} \\
		E-Mail: \{%
            \href{mailto:sterz@informatik.uni-marburg.de}{sterz},
            \href{mailto:msommer@informatik.uni-marburg.de}{msommer},
            \href{mailto:luettge@informatik.uni-marburg.de}{luettge},
            \href{mailto:freisleb@informatik.uni-marburg.de}{freisleb}%
            \}@informatik.uni-marburg.de
	}
}

\maketitle

\begin{abstract}
Recent advances in low-cost microcontrollers have enabled innovative smart home applications.
However, existing systems typically consist of single-purpose devices that only report sensed data to a controller.
Given the potential for residential emergencies, we propose to integrate emergency detection systems into smart home environments.
We present an ad-hoc distributed sensor network (DSN) designed to detect five common residential emergencies: fires, gas and water leakages, earthquakes, and intrusions.
Our novel approach combines diverse sensors with a voting-based consensus algorithm among multiple nodes, improving accuracy and reliability over traditional alert systems.
The consensus algorithm employs a majority rule with weighted votes, allowing adjustments for various scenarios.
An experimental evaluation confirms our approach's effectiveness in accurately detecting emergencies while demonstrating reliability in mitigating node failures, ensuring system longevity, and maintaining robust communication.
Additionally, our approach significantly reduces power consumption compared to alternatives.
\end{abstract}
\section{Introduction}
\label{sec:introduction}
Significant progress has been made in producing low-cost, wireless microcontrollers, paving the way for a wide range of low-power sensors for Internet of Things (IoT) scenarios.
Smart home applications are particularly interesting IoT domains, having the highest impact on peoples' life.
They range from monitoring to controlling devices like motion-based lighting or remote-controlled heating systems~\cite{astell2020s,SurveyWSN}.

An important smart home application is increasing residential safety by monitoring environmental parameters to detect potential emergencies and trigger responses.
However, such applications require social acceptance to avoid deactivation due to lack of accuracy or reliability~\cite{marikyan2019systematic,astell2020s,AcceptanceCriteriaSmartEnvironment,MultiSensorSystems,baek2021intelligent}.
In residential scenarios, a larger area must be covered, requiring distributed sensors across multiple communicating nodes.
Potential threats are often characterized by a combination of different parameters, necessitating multiple sensors on multiple nodes.
For example, fires can be distinguished from cooking in the kitchen if both a smoke sensor and a temperature sensor are used in multiple rooms.

We present a novel decentralized multi-sensor approach using multiple nodes to detect potential emergencies.
Our approach is based on a distributed sensor network (DSN) designed to sense various parameters of predefined scenarios.
We present a two-level voting algorithm to increase accuracy, i.e., sensitivity and specificity, for reliably detecting emergencies.
The first level includes using multiple sensors on a single node to improve specificity.
Using multiple nodes improves the sensitivity compared to a single-node approach.
The second level includes voting among all nodes of the network, providing an additional reliability level and decreasing false positives.
Giving a node a specific weight supports adjusting the system's performance to users' individual requirements.

Our evaluation shows that the approach can accurately detect five pre-defined events: fires, gas and water leakages, earthquakes, and intrusions, by setting appropriate detection thresholds for relevant sensors.
The system maintains high accuracy even with node failures or message loss.
Our proof-of-concept implementation can run over multiple months.
A single node per room in a residential building with appropriate communication range suffices for an accurate and reliable emergency detection system.
The source code of our implementation is released under a permissive open-source license\footnote{\url{https://github.com/umr-ds/residential_safety/}}.

\section{Related Work}
\label{sec:rel_work}
IoT studies have investigated sensor networks for environmental monitoring and disaster management, emphasizing AI-driven data analysis for improved detection and maintenance~\cite{DisasterManagementIoT,SurveyWSNUpToDate,SurveyWSN}.
Machine learning methods in DSNs play a crucial role in event detection~\cite{EventDetectionMethods}.
Various organizational structures and communication models have been explored in DSNs~\cite{gulati2022review}.
Emergency detection approaches typically use sensor-based monitoring for specific emergency types.
Multi-sensor strategies have been advocated to enhance event detection rates~\cite{baek2021intelligent,IoTModellingFire}.
Sensor fusion algorithms have shown superior detection rates and reliability~\cite{DesignAndImplementationMobileFire,savaglio2019lightweight}.
Fuzzy-based and decentralized approaches have been used to reduce power consumption and enhance efficiency~\cite{KalmanApproaches,sun2020distributed}.
Voting algorithms have improved event detection rates~\cite{sharma2020sensor}, while decentralized databases and fail-safe middleware have optimized data processing and system robustness~\cite{al2023probabilistic}.
Some methodologies focus on specific scenarios or rely on single sensors~\cite{IoTModellingFire,earthquakeMEMSaccelerometer,IntrusionDetectionSmartHome}.
Energy-efficient frameworks and communication protocols have been developed to optimize node selection and minimize energy consumption~\cite{pal2023smart,sterz2023energy}.
Existing approaches often lack distributed event detection or are specialized to single events, and some approaches rely on single nodes.
Additionally, centralized communication models pose risks of single-point of failure situations.
\section{System Design}
\label{sec:design}
Our general system design for emergency detection is based on multiple sensors on multiple communicating nodes to increase detection reliability and accuracy.
For example, intrusions are detected by a combination of hall and passive infrared (PIR) sensors.
This increases specificity, since two sensors are now required to detect a potential emergency.
Using multiple nodes increases sensitivity, ensuring that emergencies are detected even if a single node might fail.
Since there is no central controller, the nodes have to reach consensus using a robust voting algorithm.

\subsection{Emergency Detection}
\label{subsec:04:scenarios}
Our approach to emergency detection is based on thresholds, where an event is categorized as potentially relevant if the corresponding threshold of the sensors for the scenario is exceeded.
This will initiate a voting to reach consensus about whether the event is an emergency or not.

\subsection{Communication}
Nodes communicate directly with each other using a wireless protocol over a single hop for two reasons.
First, this adds locality to the system.
Although more nodes increase reliability and accuracy, covering a large area could lead to situations where an emergency could be dismissed because parts of the network are not within its area.
When only nodes in a certain proximity are voting for the emergency, sensitivity will be increased.
Second, it ensures that node failures do not affect the availability of other nodes, since message forwarding or multi-hop communication through other nodes is avoided.
Furthermore, to reduce complexity and unnecessary power usage, communication is only initiated if a node recognizes an emergency or receives a voting request.

\subsection{Reaching Consensus}
\label{design:sub:consensus}
To decide whether a detected event indicates an emergency or is a false positive, consensus must be reached between nodes.
First, the node will first compute its vote.
To do so, the sensors on the node are measured for a short amount of time, and a mean value is calculated.
If this mean value exceeds the corresponding threshold, the vote is set to the mean value and zero otherwise.
When the voting algorithm is initiated due to a vote request, the vote is sent to the initiating node.
In addition, the receiver of a voting request starts its own instance of the voting algorithm.
This cascading effect ensures redundancy if a node fails during the voting phase.
Once the initiating node receives all votes or a timer has expired, the network's decision is calculated based on the received votes.
The vote of each participating node is multiplied by its associated node weight that is sent with the voting answer message.
The node weight is used to indicate the confidence of the node.
These votes are summed up to form a total vote, which is then compared with a required majority value depending on the particular scenario.
To address reliability issues with node weights, we implemented a balancing mechanism.

\section{Implementation}
\label{sec:implementation}

\subsection{Potential Threats and Use Cases}
We consider five types of emergency in our current implementation of our system: fires, gas leaks, water leaks, earthquakes, and intrusions.

\subsubsection{Fires}
\label{subsec:04:fire}
Fires are among the most severe residential emergencies, particularly dangerous at night when residents are asleep~\cite{FireRiskFactors}.
Most fatalities are caused by smoke inhalation, especially CO~\cite{machado2021towards}.
Our approach detects changes in CO or odorized gas concentrations.
A combination of CO concentration and temperature rise has proven most specific and sensitive~\cite{baek2021intelligent}.
Our gas leakage experiments showed that the CO sensor lacks specificity.
Thus, CO and odorized gas sensors jointly indicate potential fires, with a temperature sensor discriminating between fires and gas leaks.

\subsubsection{Gas Leakages}
\label{subsec:04:gas_leakage}
Most industrial gases pose health risks if inhaled~\cite{GasLeakageRisk}.
Commonly used gases are odorized to allow detection of small concentrations~\cite{michanowicz2023natural}.
However, this is not entirely reliable due to individual differences in smell sensitivity due to diseases or age~\cite{michanowicz2023natural}.
As noted in Section \ref{subsec:04:fire}, the CO sensor can be triggered by gas leakages, necessitating the addition of a temperature sensor for discrimination.

\subsubsection{Water Leakages}
\label{subsec:04:water_pipe_burst}
Water leakages from pipe bursts can cause substantial damage~\cite{opticalWaterpipe}.
Our system employs non-invasive detection methods, since residents may not know exact pipe locations and building alterations may be infeasible.
Water probes are placed on floors or in areas where a leakage is possible, including near potential sources like dishwashers or washing machines.
Due to the lack of other non-invasive and location-independent liquid detection options, the water-level sensor is our sole detector for potential water leakages.

\subsubsection{Earthquakes}
\label{subsec:04:earthquakes}
Like fires, earthquakes are very dangerous emergency scenarios~\cite{mase2022liquefaction}.
Their threat potential ranges from weak tremors damaging furniture to strong quakes causing structural damage and potential building collapses.
Studies have shown that low-cost MEMS accelerometers provide suitable earthquake detection~\cite{earthquakeMEMSaccelerometer}\cite{cremen2020earthquake}.
We utilize an accelerometer to detect potential earthquakes.
Due to its high specificity, the accelerometer is the only sensor used for this purpose\cite{earthquakeMEMSaccelerometer}.

\subsubsection{Intrusions}
\label{subsec:04:intrusion}
Residential burglary is one of the most common theft-related crimes in Germany~\cite{psk2022}.
While standard intrusion detection often uses camera-based approaches~\cite{IntrusionDetectionSmartHome}, these are unsuitable for energy-efficient smart homes due to high power consumption and privacy concerns~\cite{IntrusionDetectionSmartHome}.
Intrusion detection is closely linked to movement or activity detection, a common use case in smart homes~\cite{ActivityDetectionPIRandHall}.
PIR sensors have proven effective for detecting human movements~\cite{IntrusionDetection,IntrusionDetectionSmartHome}.
We also use a hall sensor, since combining hall and PIR sensors effectively detects movement and determines if a person enters or leaves a room~\cite{ActivityDetectionPIRandHall}.
The hall sensor monitors the status of doors.
Since these sensors cannot differentiate between residents and intruders, intrusion detection is deactivated by default but can be user-activated, avoiding false detections from normal daily activities.

\subsection{Hardware}
\label{impl:sub:hardware}
Each node consists of a microcontroller and seven sensors.
The ESP32-PICO-MINI-02U Revision 2\footnote{\url{https://www.adafruit.com/product/5400}} microcontroller is responsible for aggregating measurements, initiating the voting algorithm, and managing the communication with other nodes.

The Adafruit LIS3DH Triple-Axis accelerometer\footnote{\url{https://www.adafruit.com/product/2809}} is used as our indicator for potential earthquakes.
An earthquake is detected when the acceleration on at least one dimension exceeds a threshold.
A MEMS accelerometer is the only sensor for earthquake detection in our approach~\cite{earthquakeMEMSaccelerometer}.

If one or both gas sensors exceed their thresholds, the Adafruit AHT20\footnote{\url{https://www.adafruit.com/product/4566}} temperature sensor distinguishes between a potential fire and gas leakage.
The gradient of a temperature rise within a defined interval is measured, improving robustness against ordinary environmental temperature changes like solar radiation or day-night cycles.
A sufficiently high gradient indicates a potential fire; otherwise, a potential gas leakage is assumed.
If gas sensors show high selectivity, the temperature sensor adds an extra level of trust to the system, since the combination of CO concentration and temperature rise has proven to be reliable for fire detection~\cite{baek2021intelligent}.

CO is a major risk and also a reliable indicator of potential fires~\cite{baek2021intelligent,machado2021towards}.
Odorized gases, in contrast, indicate a potential gas leakage that the DFRobot SEN0564\footnote{\url{https://www.dfrobot.com/product-2697.html}} CO sensor might not recognize.
In addition, the DFRobot SEN0571\footnote{\url{https://www.dfrobot.com/product-2710.html}} odorized gas sensor provides a reliable measurement for fire detection, since smoke is also recognizable as odorized gas.
The temperature sensor differentiates between CO and gas leakages due to the high cross-sensitivity of the two gas sensors.
If a potential fire is indicated by the temperature sensor, both gas sensors' values are considered.
Since gas sensors provide only qualitative measurements, a mean value of 300 measurements is calculated after preheating, serving as a reference for setting an appropriate threshold.

To detect water leakages, we place probes near potential sources of water leakages, such as dishwashers, washing machines, or underneath wash basins.
Other methods, such as infrared distance measurements or fiber optic methods, are invasive to install or error-prone due to vibrations~\cite{opticalWaterpipe,LidarWaterLevel}.
Thus, the DFRobot FS-IR02 water-level\footnote{\url{https://www.dfrobot.com/product-1470.html}} sensor is our only sensor used to detect water leakages.

To detect intrusions, we rely on a HC-SR501 PIR\footnote{\url{https://www.mpja.com/download/31227sc.pdf}} sensor, which has been widely used for this application \cite{IntrusionDetectionSmartHome, IntrusionDetection}.
If the PIR sensor detects a movement inside its field of view, it triggers the voting algorithm, which then consults the hall sensor described below.

The KY-024 Linear Magnetic Hall sensor\footnote{\url{https://cdn.shopify.com/s/files/1/1509/1638/files/Hall_Sensor_Modul_Datenblatt.pdf?11496986819545999115}} uses a magnet to observe the door status.
The sensor must be placed near door or window hinges, allowing hinge movement to move the magnet towards the hall sensor.
The sensor recognizes the magnetic field and triggers an alarm when the door opens.
When closed, the magnet moves away, and its field is no longer detected.
This design has limitations: if a door is accidentally left open and the PIR sensor falsely detects movement, the system may interpret this as a potential intrusion.

\subsection{Software Design}
Our software design employs cyclical program execution, divided into the node's uptime and a subsequent deep sleep phase.
This design reduces power consumption and allows nodes to recognize ongoing voting during uptime.

Uptime is divided into idle uptime, where the node listens for incoming vote initialization messages, and active uptime, where the node may participate in voting, including transmitting messages and reading sensor measurements.

\subsubsection{Deep Sleep Phase}
The deep sleep phase is crucial for energy efficiency and may be interrupted by external peripherals, the ultra-low power (ULP) co-processor, or a timer.
After uptime, the microcontroller initiates deep sleep, starting the ULP program and setting sensor thresholds.
An adjustable wake-up timer ensures the node can receive potential voting requests by waking up the main processor after a set period.

During deep sleep, the ULP reads sensor measurements, compares them to preset thresholds, and, if exceeded, sets the wake-up reason, executing the main program in uptime.

\subsubsection{Communication}
Communication between nodes uses the ESP-NOW~\footnote{\url{https://www.espressif.com/en/solutions/low-power-solutions/esp-now}} protocol, a low-power, connectionless peer-to-peer Wi-Fi protocol by Espressif, with a default bit rate of 1.0 Mbps.
The Long Range Wi-Fi mode achieves distances up to one kilometer with bit rates of 0.5 Mbps and 0.25 Mbps.

Three message types are used: (i) voting request, (ii) voting response, and (iii) voting notification.
The voting request starts the voting process, the voting response contains the node's vote and weight, and the voting notification informs nodes of consensus and emergency detection.

\subsubsection{Reaching Consensus}
The voting algorithm runs on the microcontroller's second core, allowing parallel communication without blocking.
Voting starts with a request message containing emergency information, followed by the node computing its vote with updated measurements.

If not all nodes have voted, the task sends the voting message to remaining nodes.
If nodes do not respond in time, vote re-balancing is initiated.
Once all votes are available, the result is calculated (\Cref{design:sub:consensus}).
A voting result message with the algorithm's decision is sent to all nodes before the task deletes itself, and the main program continues.
\section{Evaluation}
\label{sec:evaluation}

\subsection{Experimental Setup}
\label{eval:sec:setup}
For our experimental evaluation, we developed five prototypes, each consisting of an ESP32 microcontroller and seven sensors, as described in \Cref{impl:sub:hardware}.
Throughout the evaluation, the thresholds to detect a particular emergency were identified experimentally.

\subsubsection{Emergencies}
\label{eval:subsub:example}
\paragraph{Fire}
Our fire simulation was performed by placing nodes at an enclosed location like an oven.
To simulate smoke production of a fire, a small smoldering fire was caused by burning a piece of kitchen paper inside this enclosed location.
This ensures that the gas concentration induced by the small fire is not diluted too much.
As mentioned in \Cref{subsec:04:fire}, the voting algorithm uses the temperature sensor as an indicator to distinguish whether a fire or gas leakage might have occurred.
The oven was preheated to 50\textdegree C before the nodes were placed inside to provide a suitable temperature gradient for the sensor to indicate a fire.
Once all preparations were finished, an increasing number of nodes was placed inside the oven.
The node weights of the voting algorithm for each node were set to 1.0.
The vote of each node was composed of both of the gas sensors.
The required majority value for this scenario was set to 2.5, considering individual false positives.

\paragraph{Gas Leakage}
Our gas leakage simulation was also conducted inside an oven that was not preheated.
A container with acetone was placed inside the oven to simulate the gas leakage.
After that, one node at a time was placed in the oven with a defined time interval in between.
Each node's node weights were set to 1.0, and each node's vote only indicated whether the odorized gas sensor exceeded the defined threshold.
The value of the required majority was set to 2.5 to provide robustness against individual false positives.

\paragraph{Water Leakage}
Water leakages were simulated by putting the water level probe into a glass of water.
To take scenarios into account where a local water leak is likely (e.g., besides a dishwasher), the node weights were set so that the voting algorithm should agree on an emergency, even if only a minority of nodes agree on it.
In this case, the node next to the dishwasher should be given a higher priority in the voting algorithm than nodes in other areas.
Therefore, the node weight of one node was set to 2.5.
This allows the node to decide the vote on its own.

\paragraph{Earthquake}
In case of earthquakes, the entire building and, thus, all nodes experience an increased acceleration simultaneously.
This means that a majority is only reached if all available nodes exceed their thresholds.
Our simulation of an earthquake was achieved by dropping a fixed mass from a fixed height to a table with the nodes placed at that table.

\paragraph{Intrusion}
Our intrusion detection was evaluated by simulating various cases that might occur at a potential break-in.
The potentiometers at the  PIR sensors were configured so that their delay is minimal and their sensitivity was maximal, providing a detection range of seven meters.
To simulate the intrusion scenario for evaluation purposes, all nodes were placed at a desk, and their PIR sensors were fixed to face in one direction.
Some of the PIR sensors were covered, depending on the respective scenario, to simulate different scenarios with movement in various rooms.
The different door states of each room were simulated by placing or removing a magnet at or from the hall sensors.
We evaluated six cases to provide a general performance of different scenarios.
Case (i) describes the typical case where no anomalies are detected.
Case (ii) simulates malfunctioning sensors or doors that have not been closed.
Case (iii) describes a potential intrusion, since it detects movement in room 2 and an open door in room 1.
This case is marginal, since the PIR sensor might malfunction, as might the hall sensor of room 1, but it could also indicate a possible intrusion emergency in cases where burglars gain access in other ways.
Another borderline case is case (iv), where only one PIR sensor detects an emergency that could be caused either by malfunctioning or intruders that have gained access in other ways.
Case (v) shows different possible combinations that might occur when walking from one room to another, including opening doors or using already opened doors.
Case (vi) is not a realistic scenario since it requires movement in all rooms simultaneously and is only provided to show the performance of the algorithm at a larger scale.

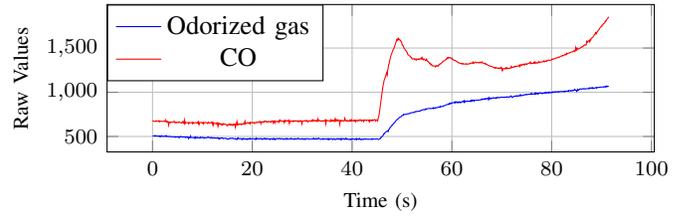
\begin{figure}
        \centering
        \begin{tikzpicture}
            \begin{axis}[
                width=\columnwidth,
                height=.4\columnwidth,
                xlabel={Time (s)},
                ylabel={Raw Values},
                grid=both,
                legend style={at={(0,1)},anchor=north west},
                tick label style={font=\footnotesize},
                label style={font=\footnotesize},
            ]
            \addplot[scatter, no marks, draw=blue]  table [x expr=\thisrow{Time}/1000, y=Odor, col sep=comma] {csv/evaluation/fire/smouldering_fire.csv};
            \addplot[scatter, no marks, draw=red]  table [x expr=\thisrow{Time}/1000, y=CO, col sep=comma] {csv/evaluation/fire/smouldering_fire.csv}; 
            \addlegendentry{Odorized gas}
            \addlegendentry{CO}
            \end{axis}
        \end{tikzpicture}
        \caption{Odorized gas and CO measurements of a smoldering fire}
        \label{fig:fire_measurements}
\end{figure}

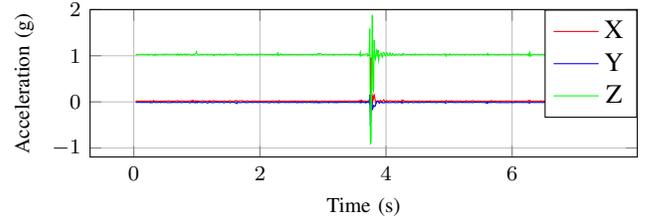
\begin{figure}[t]
    \centering
        \begin{tikzpicture}
            \begin{axis}[
                width=\columnwidth,
                height=0.4\columnwidth,
                xlabel={Time (s)},
                ylabel={Acceleration (g)},
                grid=both,
                ymax = 2,
                legend style={at={(1,1)},anchor=north east},
                tick label style={font=\footnotesize},
                label style={font=\footnotesize},
            ]
            \addplot [scatter, no marks, draw=red] table [x expr=\thisrow{Time}/1000, y=X, col sep=comma] {csv/evaluation/acceleration/1kg_40cm.csv};
            \addlegendentry{X}
            \addplot [scatter, no marks, draw=blue] table [x expr=\thisrow{Time}/1000, y=Y, col sep=comma] {csv/evaluation/acceleration/1kg_40cm.csv};
            \addlegendentry{Y}
            \addplot [scatter, no marks, draw=green] table [x expr=\thisrow{Time}/1000, y=Z, col sep=comma] {csv/evaluation/acceleration/1kg_40cm.csv};
            \addlegendentry{Z}
            \end{axis}
        \end{tikzpicture}
    \caption{Acceleration measurement of a 1 kg mass dropped from a fixed height of 40 cm}
    \label{fig:earthquake:massdropped_40cm}
\end{figure}

\subsection{Evaluation Metrics}
\subsubsection{Accuracy}
To evaluate the accuracy of our system, we performed the tests described in \Cref{eval:subsub:example} and measured whether the system is generally able to detect those emergencies.
To do so, we first performed preliminary tests to estimate reasonable thresholds for the sensors used, which were then configured for the tests.
Note that the accuracy test is a qualitative test, evaluating only whether the system is generally capable of detecting these emergencies.

\subsubsection{Reliability}
\label{eval:subsub:reliable}
In this subsection, the reliability of the system is evaluated using three aspects: node failures, power consumption, and communication range.

\paragraph{Node Failures}
Detecting emergencies despite node failures is a crucial aspect of a sensor network, since malfunctioning sensors or nodes may induce unexpected behavior to the entire system. 
The behavior of the system was investigated using the water leakage emergency as an example to evaluate reliability in terms of node failures or message losses.
After this initial evaluation, an increased number of nodes was switched off and the algorithm's results were evaluated in terms of correct voting and acceptance of an event.
An event should still be accepted if a majority of votes with the adjusted node weights vote for it.

\label{sec:07:powerconsumption}
\begin{figure}[t]
    \centering
    \includegraphics[width=.8\columnwidth]{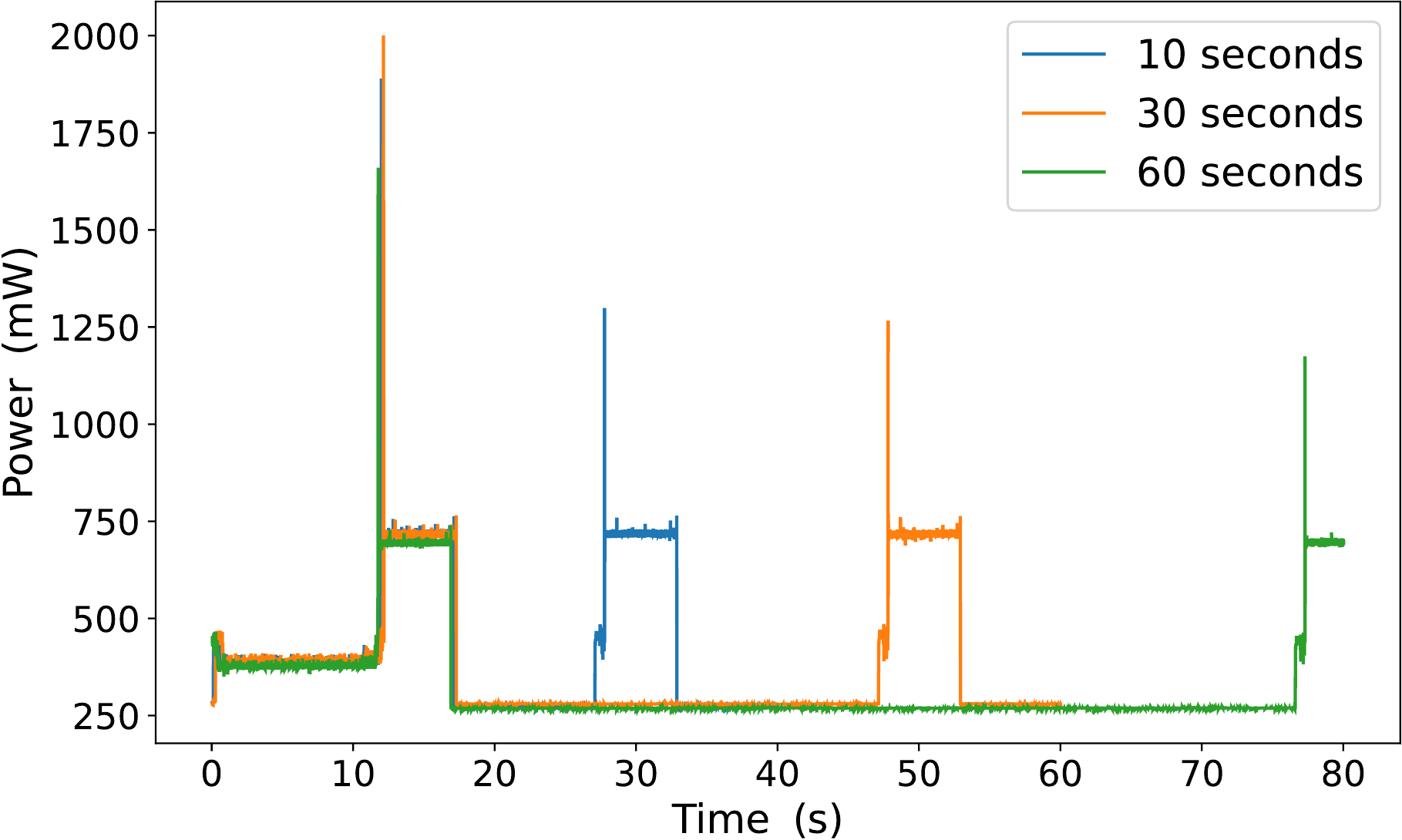}
    \caption{Power consumption of one node for different deep sleep intervals}
    \label{fig:current_consumption_single_node}
\end{figure}

\paragraph{Power Consumption}
The power consumption of a single node with three different deep sleep intervals and all its sensors was measured using a Monsoon High Voltage Power Monitor\footnote{\url{https://www.msoon.com/high-voltage-power-monitor}}.
This provides an overview of the influence of different deep sleep intervals on overall energy consumption.
The deep sleep intervals were set to 10, 30, and 60 seconds.

\paragraph{Communication Range}
A standalone application was implemented to evaluate the ESP-NOW protocol's indoor communication range.
One node transmitted 1,000 messages with dummy data to other nodes with a 10 ms delay between messages.
The receiving nodes used a counter to track received messages.
This experiment, involving one receiver and one transmitter, was repeated 50 times across various locations to simulate common living scenarios, considering walls and floors that could affect signal strength and packet loss.
Five scenarios were defined for evaluation: (i) two nodes in the same room, four meters apart; (ii) nodes six meters apart with one wall between them; (iii) nodes twelve meters apart with three walls and two doors between them; (iv) the transmitter in the basement, two floors below the receiver with several massive walls between them; (v) four transmitting nodes and one receiving node to assess interference and message loss.

\subsection{Evaluation Results}

\subsubsection{Accuracy}
\paragraph{Fire}

The red graph in \Cref{fig:fire_measurements} shows the raw values of the CO concentration measured by the CO sensor during a measurement interval of 90 seconds.
After about 30 seconds, the smoke production of a smoldering fire was simulated by burning a piece of kitchen roll, showing the used sensor's specificity.
The figure shows the reaction time between the fire initiation and increasing raw values is approximately 10 seconds.
This delay is caused by the fact that the smoke first had to disperse.
The sensitivity evaluation shows that at least three nodes are required to detect a potential smoldering fire, as described in \Cref{eval:sec:setup}.
Furthermore, since each node's vote consists of two parameters, the odorized gas, and CO values, both values need to exceed the sensor's threshold.

\paragraph{Gas Leakage}
As described in \Cref{subsec:04:fire}, the used odorized gas and CO sensors lack specificity, resulting in a cross-sensitivity between odorized gas and CO (c.f. \Cref{fig:fire_measurements}).
Nevertheless, with at least three nodes, the system is sensitive enough to detect a gas leakage.

\paragraph{Water Leakage}
Since, in this experiment, one node had a weight of 2.5, a false positive detection by this node will result in a false positive detection of the entire system.
However, this is the desired behavior in local events, such as water leakage from a dishwasher, only the node besides the dishwasher will sound an alarm and inform all other nodes.
This experiment highlights the adaptability of our system, allowing configurations for specific scenarios.

\paragraph{Earthquake}


Figure~\ref{fig:earthquake:massdropped_40cm} shows the acceleration on all three axes.
The impact of the mass is visible approximately four seconds after the start.
The peak on the z-axis reaches from -1g up to 2g.
The non-uniformity of the impact is explained by the resolution configuration, which is set to $\pm$2g and thus cuts off the peak on the z-axis at 2g.
In this scenario, four sensors could detect false positives, but the algorithm would still decline a potential emergency, showing the improved sensitivity using our proposed multi-node approach.

\paragraph{Intrusion}
This experiment underlines the low majority value on the decision of the algorithm.
In general, an event is accepted if at least two nodes recognize a potential intrusion event.
Some restrictions are emphasized in case (ii).
Case (ii) shows that although a hall sensor has recognized an open door, no voting decision has been initiated, highlighting the increased specificity of our proposed approach.

\subsubsection{Reliability}
\paragraph{Node Failures}
\begin{figure*}[t]
    \centering
    \begin{subfigure}{\columnwidth}
        \centering
        \begin{tikzpicture}
            \begin{axis}[
                width=\columnwidth,
                height=0.4\columnwidth,
                xlabel={Loop count},
                ylabel={Received Messages},
                grid=both,
                ymin = 700,
                ymax = 1100,
                tick label style={font=\footnotesize},
                label style={font=\footnotesize},
            ]
            \addplot[line width=0.5mm, no marks, draw=blue] table [x=Loopcount, y=RecvMsg, col sep=comma] {csv/evaluation/range_test/same_room.csv}; 
            \end{axis}
        \end{tikzpicture}
        \caption{Results for scenarios (i), (ii) and (iii)}
        \label{subfig:rangetest_result_scen}
    \end{subfigure}
    \begin{subfigure}{\columnwidth}
        \centering
        \begin{tikzpicture}
            \begin{axis}[
                width=\columnwidth,
                height=0.4\columnwidth,
                xlabel={Loop count},
                ylabel={Received Messages},
                grid=both,
                ymin = 700,
                ymax = 1100,
                tick label style={font=\footnotesize},
                label style={font=\footnotesize},
            ]
            \addplot[line width=0.5mm, no marks, draw=blue] table [x=Loopcount, y=RecvMsg, col sep=comma] {csv/evaluation/range_test/basement.csv}; 
            \end{axis}
        \end{tikzpicture}
        \label{subfig:rangetest_result_scen4}
        \caption{Results for scenarios (v)}
    \end{subfigure}
\caption{Comparison of successful received messages for scenarios (i), (ii), (iii) and (iv)}
\label{fig:rangetest_comparison}
\end{figure*}
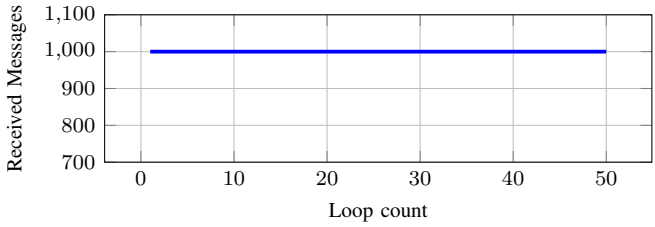
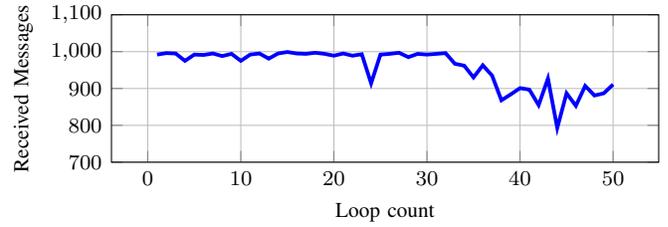

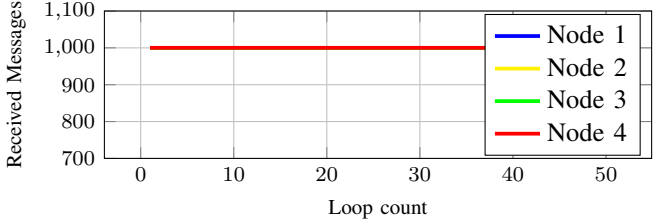
\begin{figure}[t]
    \centering
    \begin{tikzpicture}
        \begin{axis}[
                width=\columnwidth,
                height=0.4\columnwidth,
                xlabel={Loop count},
                ylabel={Received Messages},
                grid=both,
                ymin = 700,
                ymax = 1100,
                tick label style={font=\footnotesize},
                label style={font=\footnotesize},
            ]
        \addplot [line width=0.5mm, no marks, draw=blue] table [x=Loopcount, y=RecvMsg, col sep=comma] {csv/evaluation/range_test/all_nodes/node1.csv};
        \addlegendentry{Node 1}
        \addplot [line width=0.5mm, no marks, draw=yellow] table [x=Loopcount, y=RecvMsg, col sep=comma] {csv/evaluation/range_test/all_nodes/node2.csv};
        \addlegendentry{Node 2}
        \addplot [line width=0.5mm, no marks, draw=green] table [x=Loopcount, y=RecvMsg, col sep=comma] {csv/evaluation/range_test/all_nodes/node3.csv};
        \addlegendentry{Node 3}
        \addplot [line width=0.5mm, no marks, draw=red] table [x=Loopcount, y=RecvMsg, col sep=comma] {csv/evaluation/range_test/all_nodes/node4.csv};
        \addlegendentry{Node 4}
        \end{axis}
    \end{tikzpicture}
    \caption{successful received messages for scenario (v)}
    \label{fig:rangetest_result_scen5}    
\end{figure}

If the node weights are distributed equally at all nodes, a reliable result was provided if at least three nodes were available.
If two nodes were available, one could decide if an emergency occurred by itself. 
This is because the algorithm accepts an event if the vote is equal to or higher than the required majority.
This behavior can easily be avoided by only comparing if the vote is higher than the required majority and not equal.
Furthermore, the algorithm's balancing part can keep the voting shares ratio between the nodes, even if some nodes fail.

\paragraph{Power Consumption}

\Cref{fig:current_consumption_single_node} shows the power consumption of a single node.
The power consumption during the first 10 seconds after start shows the initial setup and has been measured to be between 400 mW and 450 mW.
The Wi-Fi module is activated after the initial calibration stage, resulting in a peak consumption of 2 W.
This stage is followed by a defined interval to wait for potential incoming messages for five seconds with a power consumption of about 750 mW, mainly caused by the Wi-Fi module.
The last stage is the deep sleep phase, which is initiated at three different intervals: 10, 30, and 60 seconds.
If no emergency is detected during this phase, the microcontroller wakes up after this interval has elapsed, and the cycle is repeated.
The node's power consumption over all three deep sleep intervals is approximately 250 mW.

In summary, the average power consumption of one device with a deep sleep interval of 10 seconds is 424.87 mW, for 30 seconds 378.00 mW, and for 60 seconds 327.21 mW.
If the node does not sleep at all, the average power consumption is about 719.75 mW.
With an average consumption of 327 mW at a deep sleep interval of 60 seconds, a node can operate for up to 330 hours if it is supplied from a common 9 V battery, such as those found in regular smoke detectors.
However, note power consumption depends highly on different properties.
Extending the deep sleep phase dramatically improves longevity.
For example, operating a node only in deep sleep mode with an average power consumption of about 250 mW already increases this duration up to 450 hours.
Furthermore, our system always uses all sensors for all scenarios.
However, equipping a node solely responsible for detecting earthquakes with temperature and gas sensors is questionable.
Therefore, a node only equipped with the microcontroller and the accelerometer would result in 6,000 hours of operation.
Finally, using more efficient components would reduce the power consumption even further.

\paragraph{Communication Range}
\label{sec:07:communication}

\Cref{fig:rangetest_comparison,fig:rangetest_result_scen5} show the results of the range evaluation.
Each graph shows the number of messages received at each loop for the scenarios described in \Cref{eval:subsub:reliable}.
\Cref{subfig:rangetest_result_scen} shows that all transmitted messages were successfully received through different distances and numbers of obstacles.
However, placing a node in the basement, as in scenario (iv), provided several massive walls as obstacles, which decreased the rate of successful messages received.
Although the rate of successful transmissions dropped after 35 loop iterations, caused by a closed door adding a new obstacle to the transmission path, an average of 95.7\% of all transmitted messages were received.
\Cref{fig:rangetest_result_scen5} shows the results of scenario (v).
Each node's rate of received messages remained at 1000.
This indicates that the network is capable of managing multiple transmissions simultaneously and is thus not susceptible to faults caused by signal interference.

The results of the range tests show that the system can fully function with only one node per room.
The Wi-Fi module's range is sufficient, and obstacles between nodes can be overcome.
Even over two floors, most of the messages arrive.
However, we do not recommend leaving too much space or too many obstacles between the nodes, i.e., ensuring a high node density between floors is also important.
Furthermore, a higher number of nodes might be useful for larger rooms and scenarios where the coverage area of a sensor is not sufficient to cover the entire room, but this only adds little value to the system in other scenarios in terms of accuracy.

\section{Conclusion}
\label{sec:conclusion}
We demonstrated that a multi-sensor DSN with a voting algorithm for node consensus effectively detects indoor emergencies in residential settings.
By setting appropriate sensor thresholds, the system accurately detected events like fires, gas and water leakages, earthquakes, and intrusions. It maintains high accuracy despite node failures or lost messages.
Properly configured, the implementation can run for months, with communication range and node distribution ensuring reliable detection.
Our code is released under an open-source license.

Future work includes real-world experiments, evaluating more sophisticated voting algorithms for increased reliability and accuracy, and using dynamically adjusted thresholds via machine learning to minimize manual adjustments.
Furthermore, methods to keep the microcontroller in sleep mode longer, such as external emergency wake-ups, could further reduce power consumption.
\section*{Acknowledgment}
This work is funded by the Hessian State Ministry for Higher Education, Research and the Arts (HMWK) (LOEWE emergenCITY), and the German Research Foundation (DFG, Project 210487104 - Collaborative Research Center SFB 1053 MAKI).

\bibliographystyle{IEEEtranS}
\bibliography{IEEEabrv,09_literature}

\end{document}